\documentclass[%
reprint,
superscriptaddress,
 amsmath,amssymb,
 aip,
]{revtex4-2}

\usepackage{graphicx}
\usepackage{dcolumn}
\usepackage{bm}
\usepackage{threeparttable}
\usepackage{tikz}
\newcommand*\circled[1]{\tikz[baseline=(char.base)]{
            \node[shape=circle,draw,inner sep=2pt] (char) {#1};}}

\begin{document}


\title{Femtosecond laser-induced sub-wavelength plasma inside dielectrics: II. Second-harmonic generation}
\author{Kazem Ardaneh}
\email{kazem.arrdaneh@gmail.com}
\author{Mostafa Hassan}
\author{Benoit Morel}
\author{Remi Meyer}
\author{Remo Giust}
\affiliation{FEMTO-ST Institute, Univ. Bourgogne Franche-Comt\'e, CNRS,15B avenue des Montboucons,25030, Besan\c{c}on Cedex, France}
\author{Arnaud Couairon}
\affiliation{CPHT, CNRS, Ecole Polytechnique, Institut Polytechnique de Paris, Route de Saclay, F-91128 Palaiseau, France}
\author{Guy Bonnaud} 
\affiliation{CEA, Centre de Paris-Saclay, DRF, Univ.  Paris-Saclay, 91191 Gif-sur-Yvette, France}
\author{Francois Courvoisier}
\email{francois.courvoisier@femto-st.fr}
\affiliation{FEMTO-ST Institute, Univ. Bourgogne Franche-Comt\'e, CNRS,15B avenue des Montboucons,25030, Besan\c{c}on Cedex, France}%

\date{\today}

\begin{abstract}
Second-harmonic emission at a frequency that is twice the laser frequency is an important diagnostic for nonlinear laser-plasma interaction. It is forbidden for centrosymmetric materials such as the bulk of sapphire. The symmetry, however, can be broken by dielectric discontinuities as a result of plasma generation inside a solid dielectric. In the present work, we explore the basic characteristics of experimentally observed second-harmonic emission during focusing a femtosecond Bessel beam inside sapphire. We employ three-dimensional particle-in-cell simulations and the Helmholtz wave equation for theoretical investigations. We analyze how the efficiency of second-harmonic generation and its polarization depend on the plasma parameters. We find that the second-harmonic is generated either due to the coalescence of two surface electromagnetic waves or nonlinear interaction between the transverse electromagnetic wave and the longitudinal electron plasma wave driven by linear mode conversion. Experimental results agree with the theoretical predictions and confirm the existence of over-critical plasma inside the sapphire that is essential for the resonance of plasma waves or excitation of surface plasmons.  
\end{abstract}

\maketitle

\section {Introduction}\label{Introduction}
Shaping femtosecond laser beams down to micrometer sizes allows generating so-called non-diffracting Bessel beams.\cite{Durnin87} 
A Bessel beam is a solution of the Maxwell equations in which the amplitude of the beam is described by the Bessel function of the first kind. They are useful tools for optical traps, \cite{Little04,Mcleod08} optical manipulation, \cite{Garc2002} laser particle acceleration, \cite{Hafizi97,Li05,Kumar2017} light-sheet microscopy, \cite{Fahrbach2010,Planchon2011} nonlinear optics ultrashort pulse filamentation, \cite{Gaizauskas06,Roskey07,Faccio12} and laser-material processing. \cite{Bhuyan_2010,Duocastella12,Rapp_2016,COURVOISIER2016,Razvan2018,Bergner18}

Femtosecond Bessel beams have been used in single-shot regime to create sub-wavelength voids inside transparent materials such as sapphire and fused silica.\cite{Bhuyan_2010,Froehly_2014,Rapp_2016} It was shown that tightly focusing a {microjoule} femtosecond Bessel beam inside sapphire and fused silica can create a {high aspect ratio channel} with a diameter of about 400~nm and a length of typically 20~$\rm \mu m$.  In the experiments, an absorption of more than 50 percent  of the laser energy was observed.  The absorbed energy is a few orders of magnitude greater than the required energy for creating a fully ionized plasma rod at the input laser critical density,  $\epsilon_{\rm{m}}n_{\rm c}=5.3\times10^{21}\,{\rm cm^{-3}}$ for a $\lambda_0=800$~nm laser focused inside the sapphire with the permittivity $\epsilon_{\rm{m}}=3.1$. We have demonstrated in Ref. \cite{ardaneh_2021} that, in the energy regime where voids can be opened in sapphire in single shot, over-critical plasma density is generated over several tens of micrometers in length and potentially over arbitrary distances thanks to the conical structure of the Bessel beam. \cite{Meyer2019} The plasma cross-section is elliptically shaped, with the major axis perpendicular to the laser polarization. The high absorption of typically 50 percent  was shown to be due to a collisionless mechanism. In addition, second-harmonic  generation could be detected for the first time, to the best of our knowledge, from the laser-plasma interaction {\it within} a transparent solid. We recall that second-harmonic emission is in principle forbidden from the bulk of sapphire because this material is centrosymmetric and, in the absence of  plasma, has zero second-order susceptibility $\chi^{(2)}=0$. The far-field emission pattern of the second-harmonic emission significantly depends on the plasma shape and its orientation with respect to the laser polarization.\cite{ardaneh_2021}

 \begin{figure*}[!htb]
\begin{center}
\includegraphics[width=\textwidth]{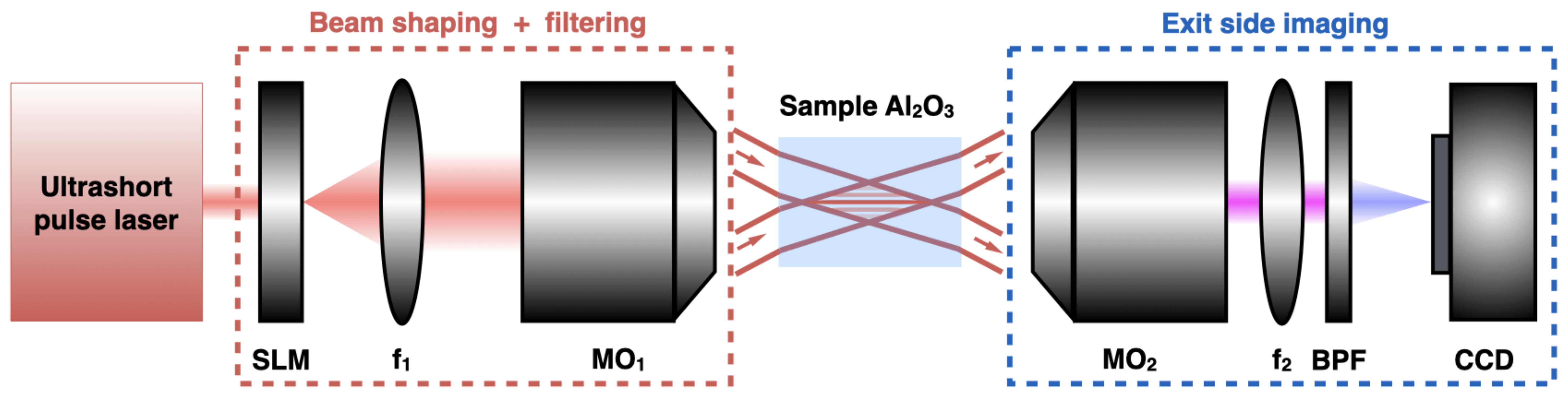}
\caption{{ Schematic of the experimental setup for femtosecond Bessel beam shaping and imaging of propagation in sapphire. Beam scanning allows reconstructing the fluence distribution inside the sample in three dimensions. Far-field scanning is achievable by changing lens f$_2$ to image the Back focal plane of the second microscope objective onto the CCD. Using a bandpass filter around $400\pm20$~nm, we select the second-harmonic component of the radiation spectrum. SLM: spatial light modulators, MO: microscope objective, BPF: bandpass filter, CCD: charge-coupled device.}}
\label{exp_setup}
\end{center}
\end{figure*}

 In the first paper of this series (Ardaneh {\it et al.},\cite{Ardaneh_2022} Paper I hereafter), we focused on the structure of the fields and the absorption mechanisms { considering two} reference PIC simulations with Gaussian and step plasma density profiles. We found that the absorption can be either due to the resonance of the electrostatic plasma waves for a transversally Gaussian density profile, or the excitation of electromagnetic surface waves in a step profile case. We also found that the particles were heated during surfing the plasma waves and the absorption process was mainly collisionless in both cases. The heated electrons were accelerated both outward and inward the plasma rod. In the outward propagation, electrostatic ambipolar fields were developed at the plasma surface due to the different mobilities for the electron and ion. On the other hand, the inward propagation led to the electron sound waves in the high-density region. 

The current paper as the second in a series is focusing on the study of the second-harmonic generation during the interaction of an ultrafast Bessel beam pulse with a plasma rod. Second-harmonic emission is conventionally an important diagnostic in the study of laser-plasma interactions. 

In an inhomogeneous plasma with density $n\gtrsim n_{\rm c}$, several mechanisms can lead to second-harmonic emission: the coalescence of two plasma waves, merging a plasma wave and an electromagnetic wave, or coalescence of two transverse waves.\cite{erokhin1969,Vinogradov_1973,Erokhin_1974,Jackel_1981,Dragila_1982} The plasma waves oscillate at roughly the laser frequency $\omega_0$ (resonance condition) and are generated either by linear mode conversion or parametric decay instability. \cite{kruer_1988,eliezer_2002,hora_2008} The second-harmonic originated from the linear mode conversion is mainly polarized parallel to the laser polarization and the spectrum is a slightly blue-shifted single peaked narrow line. \cite{Vinogradov_1973,Auer_1979, Bethune_1981,Zhizhan_1983} The parametric decay instability, however, leads to second-harmonic emission polarized perpendicular to the incident laser and the emission line is red-shifted and broadened. \cite{Vinogradov_1973,basov_1979, Jackel_1981, Zhizhan_1983}  In contrast to the blue-shifted narrow spectrum of linear mode conversion, the appearance of the red-shifted component has a  threshold for the laser intensity $\gtrsim 10^{14} \mathrm{~W/cm^{2}}$. \cite{basov_1979, Jackel_1981, Zhizhan_1983} The generated second-harmonic around the critical surface propagates in both directions. It is reflected from the high-density region implying the presence of the second-harmonic in the reflected signal.

Our paper is organized as follows. In Sec. \ref{experiments}, we first recall our experimental results and show the far-field fluence distribution of the second-harmonic, analyzed using a polarizer. We report in Sec. \ref{PIC Simulations} the results of Particle-In-Cell (PIC) simulations using two {limiting} plasma density profiles, as discussed in Paper I.  Then,   in Sec. \ref{Theory of second-harmonic generation}, we perform an analysis using the Helmholtz equation in a two-dimensional model of laser interaction with a plasma density gradient. The results from the Helmholtz equation show a good agreement with our experiments and PIC simulations, in terms of the polarization and conversion efficiency, and sheds light on the mechanisms of second-harmonic  generation at play.

\section {experiments}\label{experiments}
We spatially shaped 120~fs laser pulses at a central wavelength of 800~nm. Using a spatial light modulator combined with a 2f-2f telecentric arrangement to produce a horizontally-polarized Bessel beam with cone angle 25$^{\circ}$ in air. The pulse duration of typically 120~fs has been characterized at the sample position. After the interaction, the pulse is collected with a $\times$50 microscope objective (Olympus MPLFLN) in a 2f-2f arrangement to image the pulse on a camera (Stingray F146B, 14 bits). The imaging system, with a numerical aperture of 0.8 is placed on an independent translation stage. The far-field images were recorded on a camera using an accumulation over 500 shots when the sample is continuously translated to separate the laser impacts by 5~$\rm \mu m$. A notch filter allows for integrating on the camera the signal in a range $400\pm20$~nm. The emission could be analyzed using a polarizing cube placed at 0 or 90 degrees with respect to the pump polarization direction. { A simplified schematic of the experimental setup is illustrated in Fig. \ref{exp_setup}.}

The measurement has been performed with the Bessel beam fully inside the bulk of sapphire. We remark that the second-harmonic signal disappears when the input pulse is stretched to the ps pulse duration. { Stretching the input pulse is equivalent to reducing its intensity considering fixed pulse energy. Reducing the intensity weakens the nonlinear current source [Eq. (\ref{j2(x)})], and subsequently, reduces the radiated power for the second-harmonic.}  The signal is also reduced when part of the beam crosses one of the air-dielectric interfaces since the plasma length is reduced.

\begin{figure}[!htb]
	\centering
		\includegraphics[width=\columnwidth]{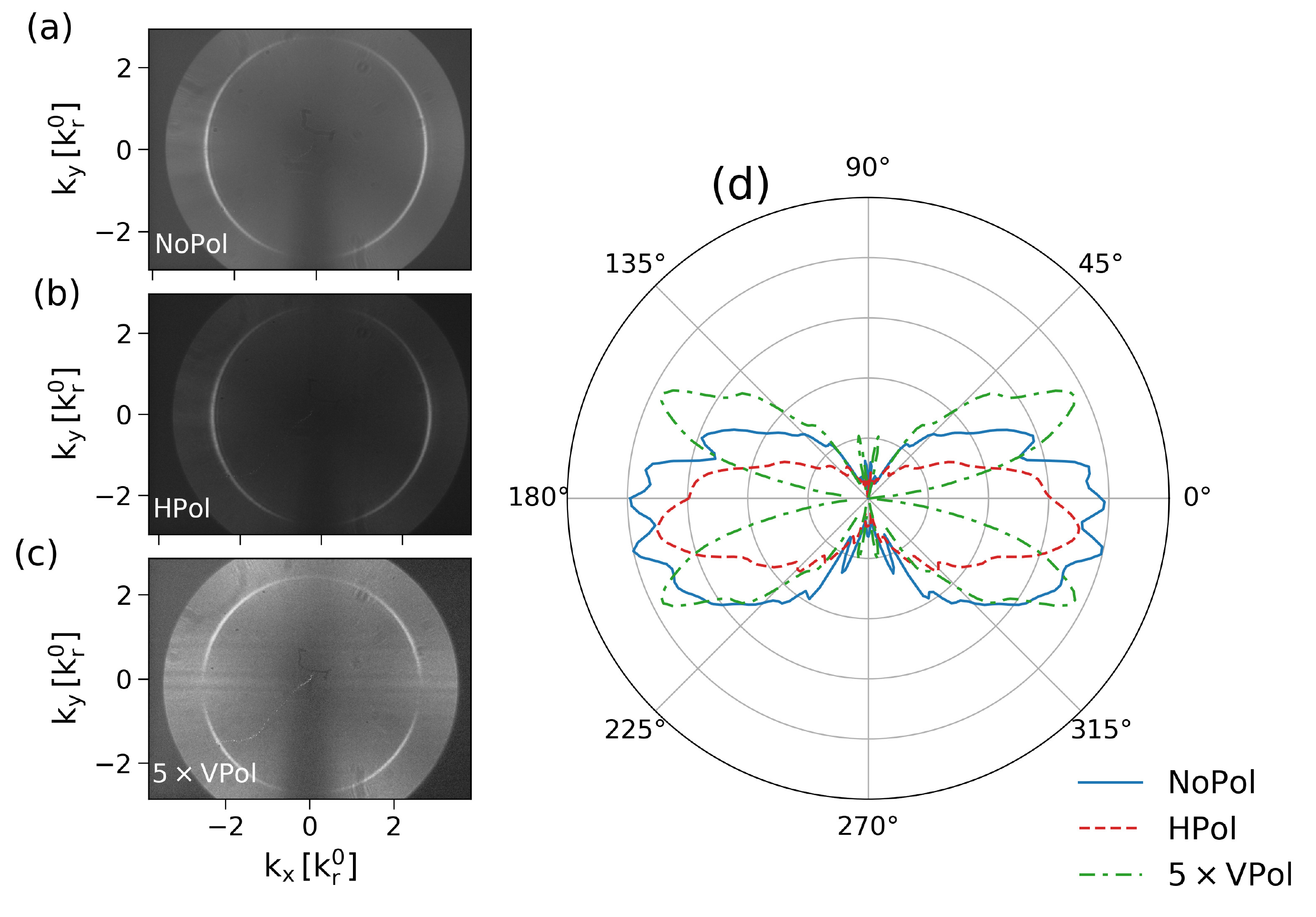}
	\caption{Far-field emission at $2\omega_0$ from the experiments: (a) without polarizer (NoPol), (b) using a horizontal polarizer (HPol), and (c) using a vertical polarizer (VPol).  The axes are normalized to $k^0_{\rm r}=k_0\sin\theta$. {The angular distribution for each distribution is shown in panel (d),  blue solid for NoPol, red dashed for HPol, and green dashed-dotted for VPol.} }
	\label{FarField}
\end{figure}

Figure (\ref{FarField}) shows our experimental results. 
Figure (\ref{FarField}a) shows two distinct signals: first, a low-intensity, nearly uniform background that we have attributed to the black-body emission of the generated hot plasma (the temperature of 10-15~eV).\cite{ardaneh_2021} This signal shows a disk limited by the numerical aperture of the system. The second-harmonic signal also shows two narrow lobes at $k_{\rm r}\approx 2 k^0_{\rm r}$, where $k^0_{\rm r}$ is the radial component of the pump wavevector. This indicates that second-harmonic  emission occurs at an angle nearly identical to the reflection of the laser. We note that a slight black mark in the range $k_{\rm x}\approx 0$ and $k_{\rm y}<0$ is a measurement artifact. It is due to the fact that part of the emission is blocked by the material modification produced by the previous laser shot. This is why, in the polar representation in Fig. (\ref{FarField}d), the far-field distributions have been symmetrized. 

\begin{figure*}[!htb]
\begin{center}
\includegraphics[width=\textwidth]{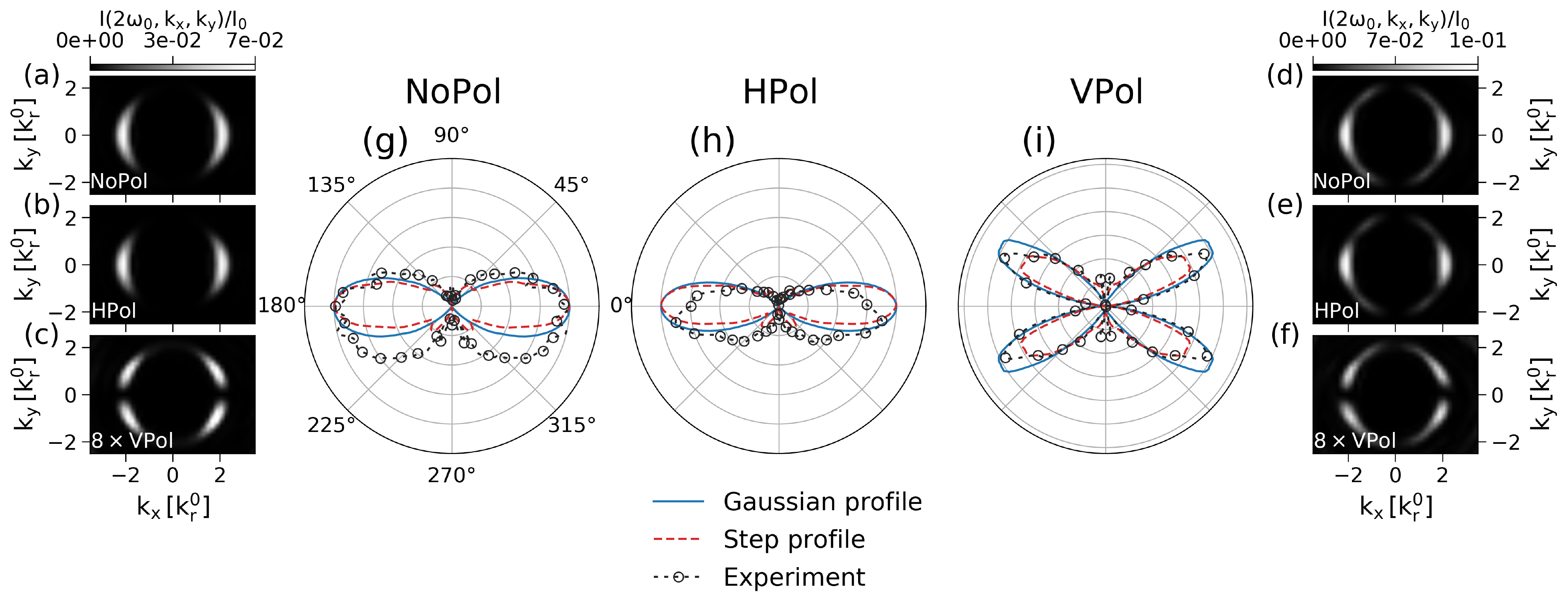}
\caption{Far-field emission at $2\omega_0$ from the PIC simulations for the Gaussian density profile (left column), and step density profile (right column).  The $2\omega_0$ emission patterns are shown: (a) and (d) for mixed polarization (NoPol), (b) and (e) for horizontal polarization [HPol, $(E_{\rm x},E_{\rm y},B_{\rm y})$ light], (c) and (f) for vertical polarization [VPol, $(E_{\rm y},B_{\rm x},B_{\rm z})$ light]. The axes are normalized to $k^0_{\rm r}=k_0\sin\theta$ of the Bessel beam. The color-bars are normalized by the average power of input pulse. The angular distributions are shown in panel (g) for NoPol, (h) for HPol, and (i) for VPol,  blue solid for Gaussian density profile, red dashed for step density profile, and black dotted with empty circles for the experiment. }
\label{shgPIC}
\end{center}
\end{figure*}

The second-harmonic in our experiments has a dominant component polarized as the input laser pulse with a dipole-shaped emission pattern, Fig. (\ref{FarField}b) and (\ref{FarField}d). This component is filtered with a horizontal polarizer (HPol). On the other hand, there is also a faint component that is polarized perpendicular to the input laser pulse (VPol) with a quadrupole-shape pattern, Fig. (\ref{FarField}c) and (\ref{FarField}d).  The average power of the second-harmonic in the experiments is calculated as  $\langle I^{2\omega_0}\rangle={1}/{2\pi} \int {\rm d}{\phi}I(2\omega_0,\phi)$.  The average power of the second-harmonic with the horizontal and vertical polarizations are respectively 80 and 20 percent of the $\langle I^{2\omega_0}\rangle$. In the following section, we will retrieve our experimental results using numerical PIC simulations.

\section {PIC Simulations} \label {PIC Simulations}
We have performed self-consistent PIC simulations using the three-dimensional  massively parallel electromagnetic code EPOCH. \cite{Arber_2015} The simulations setups and numerical scheme are described in detail in Paper I. In brief, we used the fully ionized plasmas composed of electrons and ions at temperature of $1$~eV. We considered Gaussian and step density profiles for the plasma as two {limiting} cases for the transition between volumetric plasma waves to surface waves, that can explain the characteristic high energy absorption in our experiments (see Paper I).  For the Gaussian density profile, the high absorption is due to the resonance of the volumetric plasma waves, that are excited due to the linear mode conversion. For the step density profile, the high absorption is due to surface plasmon excitation.  

In our PIC simulations, we have used the characteristic plasma scales that could reproduce our multiple experimental diagnostics (far-field, near-field, absorption), as shown in Paper I. In the step density profile, the plasma in $xy-$plane is an ellipse with uniform density $n_{\max}=5n_{\rm c}$ and minor axis $2R_{\rm x}=380$~nm along the $x-$direction, and major axis $2R_{\rm y}=900$~nm along the $y-$ direction. For the Gaussian profile, the plasma density distribution is $n =n_{\max}\exp(-x^2/w_{\rm x}^2)\exp(-y^2/w_{\rm y}^2)$, where $w_{\rm x:y}$ is the width along the $x-$ or $y-$direction with critical radius $R_{\rm x:y}$. We injected from the $z_{\min}$ boundary a linearly $x-$polarized Gaussian pulse propagating in the positive $z-$direction. We applied a phase $\phi(r)=-k_0r\sin\theta$ to the Gaussian beam to create a Bessel-Gauss beam.\cite{Ardaneh_20} The peak intensity in the Bessel zone is $6\times 10^{14}\,{\rm W/cm^{2}}$ in the absence of plasma (pulse energy 1.2~$\rm \mu J$). There are 32 particles per cell per species. We used standard FDTD schemes and triangular particle weight profiles. 

In simulations, we calculated the far-field intensity for the second-harmonic as follows. After recording the fields at a fixed propagation distance of $z=20$~$\rm \mu m$, in a spatial window for $\vert x \vert \leqslant 5$~$\rm \mu m$ and $\vert y \vert \leqslant 5$~$\rm \mu m$,  we calculated the intensity spectrum $I(\omega,k_{\rm x},k_{\rm y})$ by performing a discrete Fourier transform on each component of the magnetic field, $B_{\rm x:y:z}(t,x,y)$. We then obtained the second-harmonic emission spectra by filtering the intensity spectrum at the frequency $2\omega_0$. 

The second-harmonic emissions from the PIC simulations are shown in Fig. (\ref{shgPIC}), for the Gaussian density profile (left side), and the step density profile case (right side). Similar to the experiments, we decompose the second-harmonic emission into horizontal and vertical polarizations relative to the $x-$polarized input laser pulse, $B_{\rm y}$ for a horizontally polarized light and $(B_{\rm x},B_{\rm z})$ for a vertically polarized light. The emission patterns are in good agreement with our experiments. One can see that the component with the horizontal polarization dominates the emission pattern. Moreover, the emission angle of the second-harmonic is $\theta^{2\omega_0}=\theta$ and therefore $k^{2\omega_0}_{\rm x:y}=2k_0\sin\theta$. {{The angular distributions for different polarizations are shown in Figs. (\ref{shgPIC}g)-(\ref{shgPIC}i). We have also included the experiment angular distributions to better visualize the agreement between simulations and experiments. The angular distribution for the Gaussian and step density profiles are the same because of the similar nonlinear current sources. For both cases, the current of heated electrons is distributed along the plasma rod and mainly parallel to the $x-$direction (see Sec. VB in Paper I)}}. 

The average power of the second-harmonic in the simulation is calculated from  $\langle I^{2\omega_0}\rangle={1}/{\Delta k_{\rm x}\Delta k_{\rm y}} \int {\rm d^2}{\bf k}I(2\omega_0,k_{\rm x},k_{\rm y})$. The conversion efficiency, {\it i.e.}, the average power of second-harmonic to the average power of input pulse $\langle I^{2\omega_0}\rangle/I_0$, is $6\times10^{-6}$ for the Gaussian density profile, and $1.5\times10^{-5}$ for step one.  The intensity of the second-harmonic is therefore on the order of $\langle I^{2\omega_0}\rangle\sim 10^{9}-10^{10}\,{\rm W/cm^2}$. 
The second-harmonic generation is more efficient in the step density profile by a factor of 2.5. { We interpret this difference in efficiency as follows. The surface waves have a larger amplitude than volumetric waves because a two-dimensional system has fewer degrees of freedom for the motion of particles. The particles are then more easily trapped and waves saturate at a larger amplitude.} 

For the Gaussian density profile, the spatially-integrated power of the second-harmonic for the horizontal and vertical components are respectively 80 and 20 percent of the $\langle I^{2\omega_0}\rangle$. These ratios are 77 and 23 percent for the step density profile. The ratio of the vertically polarized component to the horizontally polarized one is $\approx\frac{1}{4}$ which is in very good agreement with the experiments in Fig. \ref{FarField}.

\section {Theory of second-harmonic generation} \label{Theory of second-harmonic generation}
In this section, we analyze the second-harmonic  generation using a simple two-dimensional model. We start with the Helmholtz equation for the second-order magnetic field $\mathbf{B}^{(2)}$ [See Eq. (\ref{B2wave3}) in Appendix \ref{Helmholtz equation for B2}]: 

\begin{equation} \label{hlm2}
\begin{split}
\boldsymbol\nabla^{2} \mathbf{B}^{(2)}+&\frac{4\omega_0^{2}}{c^{2}}  \epsilon^{(2)}\mathbf{B}^{(2)}+\frac{1}{\epsilon^{(2)}}\boldsymbol\nabla \epsilon^{(2)} \times \left(\boldsymbol\nabla \times \mathbf{B}^{(2)}\right) \\
&=\frac{4\pi}{c}\left( -\boldsymbol\nabla \times \mathbf{J}^{(2\omega_0)} +\frac{1}{\epsilon^{(2)}}\boldsymbol\nabla \epsilon^{(2)} \times \mathbf{J}^{(2\omega_0)} \right)  \\
&=\mathbf{F}^{(2\omega_0)}
\end{split}
\end{equation}
here $\epsilon^{(2)}=\epsilon(2\omega_0)$ is the plasma permittivity at the second-harmonic frequency. We used the Drude model for the plasma permittivity $\epsilon$. Hence

\begin{align}\label{eps_plasma}
\epsilon(\omega)=1-\frac{\omega_{\rm pe}^{2}/\omega^{2}}{1+j \nu_{\rm eff}/ \omega}
\end{align}
where $\omega_{\rm pe}=(4\pi n e^2/m)^{1/2}$ is the electron plasma frequency, and $\nu_{\rm eff}$ is the effective damping frequency in the plasma. Here, we have chosen to model a plasma placed in vacuum as in the EPOCH simulations (in sapphire, $\epsilon$ and $\omega_{\rm pe}$ would be modified because of the permittivity of the solid dielectric). 

The term $\mathbf{F}^{(2\omega_0)}$ in Eq. (\ref{hlm2}) is the nonlinear source of the second-harmonic. One can obtain the associated second-order current density from the equation of motion and continuity equation for electrons coupled with the Maxwell equations. \cite{shen_2003} It reads: 

\begin{equation}\label{j2(x)}
\mathbf{J}^{(2\omega_0)}=-\frac{j n e^{3}}{4 m^{2} \omega_0^{3}} \left[4\frac{\boldsymbol\nabla\ln n \cdot \mathbf{E}}{\epsilon}\mathbf{E}+\boldsymbol\nabla(\mathbf{E} \cdot \mathbf{E})\right]
\end{equation}

The polarization of the second-harmonic $\partial\mathbf{P}^{(2\omega_0)}/\partial t=\mathbf{J}^{(2\omega_0)}$ is given by: 

\begin{equation}\label{p2omega}
	\mathbf{P}^{(2\omega_0)}=\chi^{(2)}_{\rm fe}\left[2\frac{\boldsymbol{\nabla}\ln n.\mathbf{E}}{\epsilon}\mathbf{E}+\frac{1}{2}\boldsymbol\nabla(\mathbf{E} \cdot \mathbf{E})\right]
\end{equation}
where $\chi^{(2)}_{\rm fe}={ne^{3}}/{4 m^{2} \omega_0^{4}}$ is the second-order susceptibility for free electrons. \cite{Bethune_1981} 
For a radial density profile, 
the polarization of the second-harmonic coming from the first term  is $\mathbf{P}^{(2\omega_0)}\propto \mathbf{E}(\mathbf{r}.\mathbf{E})$. Therefore, in the case of a linearly polarized laser, the emission at the second-harmonic will have parallel polarization relative to the input laser. This term vanishes for $s-$polarized light because $\mathbf{E}$ is perpendicular to the density gradient ($\mathbf{r}.\mathbf{E}=0$). The emission pattern due to this term will be two parallel lobes oriented along with the directions $\pm\mathbf{E}$ of the input laser, following a $\cos^2\phi$ in which $\phi$ is the angle between $\mathbf{r}$ and $\mathbf{E}$. The second term in the $\mathbf{P}^{(2\omega_0)}$ expression results in a $p-$polarized second-harmonic, irrespective of the polarization of the input laser.  As a result, the $s-$polarized second-harmonic is possible only for an input laser with both $s-$ and $p-$ polarization.\cite{von_1992}

We now analyze more quantitatively the influence of the plasma density gradient on the different terms in a two-dimensional model. Let us consider a plasma slab in the $xy-$plane and infinite in the $z-$direction. We assume that the plasma density linearly increases with the $x-$coordinate ($n/n_{\rm c}=1+x/L$, the origin of coordinates is chosen so that with $x>0$ $\epsilon(x)<0$, and with $x<0$ $\epsilon(x)>0$). A mixed polarized monochromatic wave [$(E_{\rm x},E_{\rm y},B_{\rm z})$ for $p-$polarization and $(E_{\rm z},B_{\rm x},B_{\rm y})$ for $s-$polarization] interacts with this plasma. The wavevector is given by ${\mathbf{k}}_{0}={\mathbf{\hat{x}}} k_{\rm 0}\cos i + {\mathbf{\hat{y}}} k_{\rm 0}\sin i$ where $i$ is the incident angle measured relative to the density gradient. 
In this geometry, $E_{\rm x}$ is the component of the electric field that drives electron plasma waves. For this geometry, the second-order magnetic field components are given by: 

\begin{equation}
\begin{bmatrix}
B^{(2)}_{\rm x}\\
B^{(2)}_{\rm y}\\
B^{(2)}_{\rm z}
\end{bmatrix}
=
\begin{bmatrix}
B^{(2)}_{\rm x}(x)\\
B^{(2)}_{\rm y}(x)\\
B^{(2)}_{\rm z}(x)
\end{bmatrix}
\times\exp{\left[2j \left(-\omega_0 t+k_{\rm y}y\right)\right]}
\end{equation}

One can obtain the source term components of $\mathbf{F}^{(2\omega_0)}$ analytically after some straightforward but lengthy algebra, see for example Ref. \cite{erokhin1969,Erokhin_1974} It reads:

\begin{equation} \label{f2(x)}
F_{\rm x}^{(2\omega_0)}=\frac{j4\pi e^{3} \epsilon^{(2)}}{m^{2} \omega_0^{3} \mathrm{c}} \left(\overbrace{\frac{2 j k_{\rm y} E_{\rm x}E_{\rm z}}{\epsilon \epsilon^{(2)}} \frac{\partial n}{\partial x}}^{\circled{2}}\right)
\end{equation}

\begin{equation} \label{f2(y)}
F_{\rm y}^{(2\omega_0)}=\frac{j4\pi e^{3} \epsilon^{(2)}}{m^{2} \omega_0^{3} \mathrm{c}} \left[-\overbrace{\frac{\partial}{\partial x}\left(\frac{E_{\rm x} E_{\rm z}}{\epsilon \epsilon^{(2)}} \frac{\partial n}{\partial x}\right)}^{\circled{2}}\right]
\end{equation}

\begin{equation} \label{f2(z)}
\begin{split}
F_{\rm z}^{(2\omega_0)}=\frac{j4\pi e^{3} \epsilon^{(2)}}{m^{2} \omega_0^{3} \mathrm{c}} &\left[-\overbrace{\frac{2 j k_{\rm y} E_{\rm x}^{2}}{\epsilon \epsilon^{(2)}} \frac{\partial n}{\partial x}}^{\circled{1}} \right. \\
&\left.+\overbrace{\frac{\partial}{\partial x}\left(\frac{E_{\rm x} E_{\rm y}}{\epsilon \epsilon^{(2)}} \frac{\partial n}{\partial x}\right)}^{\circled{2} } \right. \\
&\left.+\overbrace{\frac{j k_{\rm y}E^{2}}{2} \frac{\partial}{\partial x} \left(\frac{n}{\epsilon^{(2)}}\right)}^{\circled{3}}\right]
\end{split}
\end{equation}
where $E^2 = (E_{\rm x}^{2}+E_{\rm y}^{2}+E_{\rm z}^{2})$. One notes that second-harmonic source $\mathbf{F}^{(2\omega_0)}$ is proportional to the density gradient and vanishes in homogeneous plasmas where $\boldsymbol{\nabla} n=0$. By using the same terminology as Nazarenko {\it et al.},\cite{Nazarenko_1995} the second-harmonic generation is either due to the nonlinear interaction of two electron plasma waves, term $\circled{1}$ in Eq. (\ref{f2(z)}), or an electron plasma wave with an electromagnetic wave, term $\circled{2}$ in Eqs. (\ref{f2(x)})-(\ref{f2(z)}), or coalescence of two electromagnetic waves in an inhomogeneous plasma, term  $\circled{3}$ in Eq. (\ref{f2(z)}).  The terms $\circled{1}$ and $\circled{2}$ arise from the first term of Eq. (\ref{j2(x)}) in which electron plasma waves are at resonance where $\epsilon=0$. The term $\circled{3}$ originates from the second term of Eq. (\ref{j2(x)}) and is due to the laser radiation force density $f_{\rm RF}=(\epsilon-1)/{8 \pi} \boldsymbol{\nabla} E^{2}$ which is faint at the considered intensity.  In a homogeneous plasma, this term is purely irrotational because 
$\boldsymbol{\nabla} \times(\boldsymbol{\nabla} E^2)=0$ and can radiate just at boundaries where $\boldsymbol{\nabla} n\neq0$. \cite{Bethune_1981} For a pure $p-$polarized laser, the $F_{\rm x}^{(2\omega_0)}$ and $F_{\rm y}^{(2\omega_0)}$ components vanish while all three terms in the $F_{\rm z}^{(2\omega_0)}$ will be present. In contrast, for the pure $s-$polarized laser, the terms $\circled{1}$ and $\circled{2}$ in $F_{\rm z}^{(2\omega_0)}$ vanish.

\begin{figure}
\begin{center}
\includegraphics[width=\columnwidth]{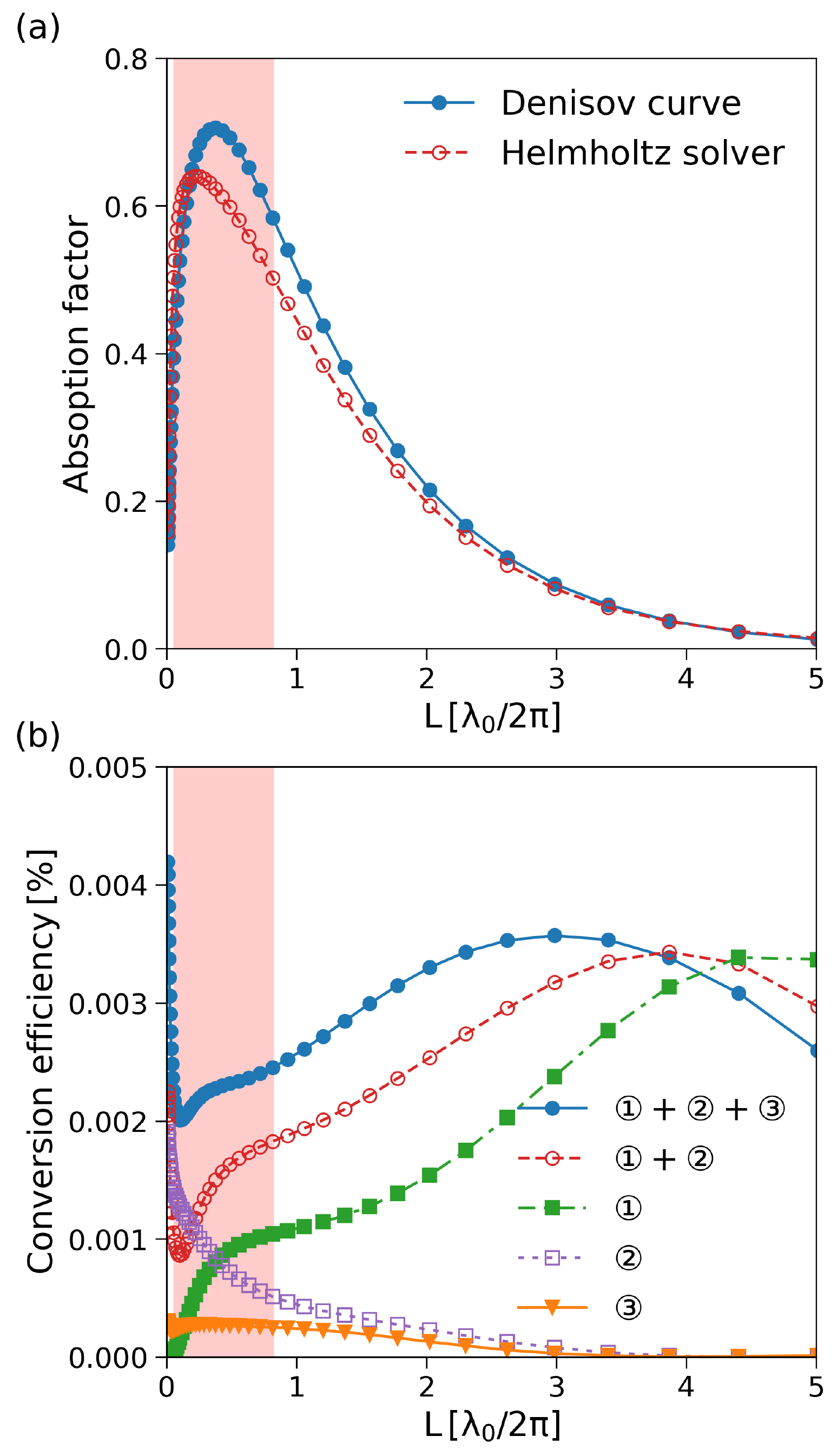}
\caption{Solutions of the Helmholtz equation for the first, and second-harmonic in the case of $p-$polarized light [nonlinear source given by Eq. (\ref{f2(z)})]: (a) absorption factor as a function of plasma scale length from the Helmholtz equation (red dashed with empty circles) versus the Denisov absorption curve (blue solid with filled circles); (b) conversion efficiency for the second-harmonic using the source term with full expression (solid blue with filled circles), the first and second terms (red dashed with empty circles), the first term (green dashed-dotted with filled squares), second term (purple dotted with empty squares), and third term (orange solid with filled triangles).}
\label{shgHH}
\end{center}
\end{figure}

The situation in the PIC simulations and experiments is as follows. In the absence of plasma, the electric field components of the Bessel beam are $E_{\rm z}\approx 0.1E_{\rm x}$ and $E_{\rm y}\approx 0.03E_{\rm x}$. However, in the presence of an elliptical plasma in the $xy-$plane, the $E_{\rm x}$, $E_{\rm y}$ components are significantly amplified and $E_{\rm y}\approx 0.5E_{\rm x}$ and $E_{\rm z}\approx 0.2E_{\rm x}$. Therefore, the dominant electric field components in the presence of the plasma are $E_{\rm x}$ and $E_{\rm y}$. This reduces the physical situation more like a $p-$polarized light with the dominant second-harmonic source given by Eq. (\ref{f2(z)}). This term generates the second-harmonic with the horizontal polarization. The vertical polarization term can be inferred from term $\circled{2}$ and the importance of the $E_{\rm z}$ relative to $E_{\rm y}$.   

 We solved the Helmholtz equations for the first-harmonic to obtain the nonlinear source and then for second-harmonic. We used an angle of incidence $i=65^{\circ}$, which corresponds to the same physical situation as in the experiment (since the cone angle $\theta=\pi/2 -i$ is the angle made with the optical axis). The solver is detailed in the Appendix \ref{Helmholtz equation solver} and is limited to the $p-$polarized case. 
 
 In Fig. (\ref{shgHH}a) we show  the absorption factor calculated from the Helmholtz equation (red dashed with empty circles) versus the Denisov absorption curve (blue solid with filled circles) as a reference for validation of our solver.\cite{Denisov_1957} The Denisov curve is accurate for slowly varying density and there is a very good agreement between the two curves in the region $Lk>3$. The deviation between the two curves is in the region $Lk<2$ where the Denisov curve is less accurate.  We show with the red rectangle the range of plasma scale length where the absorption exceeds 0.5, which corresponds to our experimental regime, as discussed in detail in Paper I.

In Fig. (\ref{shgHH}b), one can see the conversion efficiency for the second-harmonic considering different terms of the nonlinear sources $F_{\rm z}^{(2\omega_0)}$ given in Eq. (\ref{f2(z)}): the full source expression, terms $\circled{1}+\circled{2}$, term $\circled{1}$ for the coalescence of two longitudinal plasma waves, term $\circled{2}$ forthe coalescence of a longitudinal plasma wave with a transverse electromagnetic wave, and term $\circled{3}$ for the coalescence of two electromagnetic waves. 

As shown in Figs. (\ref{shgHH}), there is an optimal plasma scale length for the absorption and the second-harmonic generation. { A scale length shorter than the optimum shifts the maxima of absorption and second-harmonic generation to higher incident angles. In this configuration, the evanesce field is more confined near the turning point, weakening the plasma wave's excitation. Hence, the matching between the two plasma waves or plasma wave and transverse electromagnetic wave results in a lower efficiency for the second-harmonic generation. The plasma wave evolves into the surface wave at the limit $L\rightarrow 0$ which corresponds to the step-density profile. In this case, the angle of incidence is the sensitive parameter for efficient surface plasmon excitation and second-harmonic generation, in contrast with the Gaussian density profile where there is also an optimal scale length for efficient matching.}

 This simple model allows us to explain several of our simulations and experimental results.  (i)  We see that in the range of plasma scale lengths corresponding to our experimental case (red rectangle), the conversion efficiency of the input laser power into the second-harmonic is $\sim 1\times10^{-5}$ as we obtained in the PIC simulations. We also see that the conversion efficiency for step density profile (limit case where $L$ tends to 0) is almost twice the inhomogeneous density profile. This is also in agreement with the results from PIC simulation in Fig. (\ref{shgPIC}) where the ratio is 2.5. (ii) The ratio between the generated second-harmonic from the source term $\circled{2}$ (purple dotted with empty squares) and the full source term (solid blue with filled circles) is always $< \frac{1}{2}$. Considering that $E_{\rm z}/E_{\rm y}\approx 0.5$, the ratio between the second-harmonic  generation with vertical polarization to the horizontal polarization is $< \frac{1}{4}$; (iii) reducing the plasma scale length  reduces the probability of two plasma waves coalescence (green line with filled squares) while it increases the probabilities of a plasma wave merging with an electromagnetic wave and merging of two electromagnetic waves (orange line with filled triangles); (iv) In the range $Lk\lesssim1$ (the corresponding range for experimental absorption), the most likely process is plasma wave merging with an electromagnetic wave (purple line with empty squares); and (vi) for a step density profile, in the limit $L\rightarrow 0$, the total conversion efficiency of second-harmonic generation increases again (solid blue curve). In this limit case, plasma waves evolve into surface waves and the efficiency of term $\circled{2}$ is now due to efficient coalescence of two surface waves. 

In the limit $L\rightarrow 0$, we can easily retrieve the direction of emission of second-harmonic.  From the conservation of energy and momentum, we have

\begin{equation}\label{SHGSP}
2 \hbar \frac{\omega_0}{c}\left[\epsilon(\omega_0)\right]^{1 / 2} \sin i=\hbar \frac{2 \omega_0}{c}\left[\epsilon(2 \omega_0)\right]^{1 / 2} \sin i_{2\omega_0}
\end{equation}
Therefore, the angle of emission for the second-harmonic reads:
\begin{equation}\label{SHGSP_2}
\sin i_{2\omega_0}=\left[\epsilon_2(\omega_0) / \epsilon(2 \omega_0)\right]^{1 / 2} \sin i
\end{equation}
The second-harmonic emission by merging two surface waves might not be exactly at the same angle as the reflected pump. In PIC simulations, the emission angle of the second-harmonic is $i_{2\omega_0}=i$ and therefore $k^{2\omega_0}_{\rm x:y}=2k_0\sin\theta$ because $\left[\epsilon(\omega_0) / \epsilon(2 \omega_0)\right]^{1 / 2} = 1$ for vacuum. For sapphire, the emission angle of the second-harmonic is $i_{2\omega_0}\approx i$ because $\left[\epsilon(\omega_0) / \epsilon(2 \omega_0)\right]^{1 / 2} \approx 1$.

In regards to the experiments and PIC simulations, the component with horizontal polarization mainly originates from the resonant first term in Eq. (\ref{j2(x)}).  As discussed before, the horizontally polarized second-harmonic might therefore be due to the merging of two plasma waves, term $\circled{1}$, or merging of a plasma wave with an electromagnetic wave, term $\circled{2}$.  In the range of plasma scale length for an absorption $\gtrsim0.5$ [Fig. (\ref{shgHH})], the latter process becomes more efficient as the plasma scale length decreases.  The faint component with vertical polarization comes from the second term in Eq. (\ref{j2(x)}) and is due to the coalescence of two electromagnetic waves, term $\circled{3}$. 

\section{Conclusions}\label{Conclusions}
In this paper, we extended our investigation of femtosecond Bessel beam-induced plasmas inside the sapphire.  {A single-shot Bessel beam can create a high aspect ratio over-critical plasmas inside the dielectric, a promising medium for the second-harmonic generation due to the currents of the resonance driven hot electrons. This is not feasible with Gaussian beam because the generated plasma reflects the most energetic part of the beam and the plasma stays sub-critical.\cite{ardaneh_2021}}

Using the Helmholtz equation for the monochromatic wave, we have studied the polarization and the conversion efficiency of the second-harmonic considering different possible nonlinear interactions. Our model reproduces correctly all main characteristics of the experimental and PIC simulation results. We showed that both linear mode conversion and resonance of the surface wave generate a second-harmonic emission polarized horizontal to the input laser. The conversion efficiency from our simple model and PIC simulations is on the order of $10^{-5}$. Under the linear mode conversion and the plasma scale length corresponding to the measured absorption, the second-harmonic emission is mainly due to the nonlinear interaction between the longitudinal electron plasma wave and transverse laser wave. On the other hand, the coalescence of two surface waves leads to the second-harmonic emission in the step density profile with almost twice higher conversion efficiency than the linear mode conversion scenario.  

{
In this work, the pattern of the second-harmonic is a mark that over-critical plasma density is reached within sapphire. While it does not allow us to discriminate between homogeneous or inhomogeneous plasma profiles, the second-harmonic diagnostic can still be useful to gain insights into the parameters of the laser-generated plasma and the absorption mechanism. From a more general point of view, the epsilon-near-zero surface associated with the critical surface of the plasma rod can represent a new platform for nonlinear frequency conversion within dielectrics.}

{The analysis presented in the current work can also shed light on experiments in other centrosymmetric materials. The preliminary measurements for the fused silica showed some differences in the second-harmonic patterns with the current work. The differences are likely due to the regime of the nonlinear ionization and the resulting density profile.}

\begin{acknowledgments}
We thank the EPOCH support team for help  \url {https://cfsa-pmw.warwick.ac.uk}, and French RENATECH network. The authors acknowledge the financial supports of: European Research Council (ERC) 682032-PULSAR, Region Bourgogne-Franche-Comte and Agence Nationale de la Recherche (EQUIPEX+ SMARTLIGHT platform ANR-21-ESRE-0040), Labex ACTION ANR-11-LABX-0001-01, I-SITE BFC project (contract ANR-15-IDEX-0003), and the EIPHI Graduate School ANR-17-EURE-0002.
This work was granted access to the PRACE HPC resources MARCONI-KNL,  MARCONI-M100, and GALILEO at CINECA, Casalecchio di Reno, Italy, under the Project "PULSARPIC" (PRA19\_4980), PRACE HPC resource Joliot-Curie Rome at TGCC, CEA, France under the Project "PULSARPIC" (RA5614), HPC resource Joliot-Curie Rome/SKL/KNL at TGCC, CEA, France under the projects A0070511001 and A0090511001, and M\'{e}socentre de Calcul de Franche-Comt\'{e}.  
\end{acknowledgments}

\appendix
\section{Helmholtz equation for $\mathbf{B}^{(2)}$}\label{Helmholtz equation for B2}

The second-order current density is given by

\begin{equation}\label{Jfull2}
\mathbf{J}^{(2)}=\sigma^{(2)} \mathbf{E}^{(2)}+\mathbf{J}^{(2\omega_0)}
\end{equation}
The first term is the linear response of the medium following the Ohm law, and the second term is the second-harmonic nonlinear source. To develop wave equations for the second-order magnetic fields, let us consider Faraday's law and Ampere's law, which become

\begin{subequations}\label{Faraday_Ampere}
\begin{align}
\boldsymbol\nabla \times \mathbf{E}^{(2)}=&j\frac{2\omega_0}{c}\mathbf{B}^{(2)} \\
\boldsymbol\nabla \times \mathbf{B}^{(2)}=&\frac{4 \pi}{c} \sigma^{(2)} \mathbf{E}^{(2)}-j \frac{2\omega_0}{c} \mathbf{E}^{(2)} +\frac{4 \pi}{c} \mathbf{J}^{(2\omega_0)}
\end{align}
\end{subequations}
Substituting for $\sigma^{(2)}=\sigma(2\omega_0)=j \omega_{\rm {pe }}^{2} / 8 \pi\omega_0$ into Eq. (\ref{Faraday_Ampere}b) gives

\begin{equation}\label{Ampere2}
\boldsymbol\nabla \times \mathbf{B}^{(2)}=-j\frac{2\omega_0}{c} \epsilon^{(2)} \mathbf{E}^{(2)} +\frac{4 \pi}{c} \mathbf{J}^{(2\omega_0)}
\end{equation}
where $ \epsilon^{(2)}=1-\omega_{\text {pe }}^{2} /4\omega_0^{2}$ defines the second-order dielectric function of the plasma. Taking the curl of Eq. (\ref{Ampere2}) gives

\begin{equation}\label{B2wave}
\boldsymbol\nabla \times(\boldsymbol\nabla \times \mathbf{B}^{(2)})=-j\frac{2\omega_0}{c} \boldsymbol\nabla \times(\epsilon^{(2)} \mathbf{E}^{(2)}) + \frac{4 \pi}{c}\boldsymbol\nabla \times \mathbf{J}^{(2\omega_0)}
\end{equation}
since 
\begin{equation}\label{curllaw}
\boldsymbol\nabla \times \epsilon^{(2)} \mathbf{E}^{(2)}=\epsilon^{(2)} \boldsymbol\nabla \times \mathbf{E}^{(2)}+\boldsymbol\nabla \epsilon^{(2)} \times \mathbf{E}^{(2)}
\end{equation}
Substituting from Eq. (\ref{Faraday_Ampere}a) and Eq. (\ref{Ampere2}) for the first and second terms in right side of Eq. (\ref{curllaw}), we obtain

\begin{equation}\label{B2wave2}
\begin{split}
-j\frac{2\omega_0}{c} \boldsymbol\nabla &\times(\epsilon^{(2)} \mathbf{E}^{(2)}) = \frac{4\omega_0^{2}}{c^{2}} \epsilon^{(2)} \mathbf{B}^{(2)}\\
&+\frac{1}{\epsilon^{(2)}} \boldsymbol\nabla \epsilon^{(2)} \times(\boldsymbol\nabla \times \mathbf{B}^{(2)}- \frac{4 \pi}{c} \mathbf{J}^{(2\omega_0)})
\end{split}
\end{equation}
Substituting Eq. (\ref{B2wave2}) into Eq. (\ref{B2wave}), we obtain

\begin{equation}\label{B2wave3}
\begin{split}
\boldsymbol\nabla^{2} \mathbf{B}^{(2)}&+\frac{4\omega_0^{2}}{c^{2}} \epsilon^{(2)} \mathbf{B}^{(2)}+\frac{1}{\epsilon^{(2)}} \boldsymbol\nabla \epsilon^{(2)} \times(\boldsymbol\nabla \times \mathbf{B}^{(2)})\\
&=\frac{4 \pi}{c}\left(-\boldsymbol\nabla \times \mathbf{J}^{(2\omega_0)} +  \frac{1}{\epsilon^{(2)}} \boldsymbol\nabla \epsilon^{(2)} \times\mathbf{J}^{(2\omega_0)}\right)\\
&=\mathbf{F}^{(2\omega_0)}
\end{split}
\end{equation}

\section{Helmholtz equation solver}\label{Helmholtz equation solver}
For $p$-polarized light, it is simpler to solve the Helmholtz equation for the magnetic field $B^{{\rm (n)}}_{\rm z}$ first ($n=1$ for first-harmonic and $n=2$ for second-harmonic), and then derive the electric fields from its solution.  The solver is developed based on the presentation in Ref. \cite{gibbon_2005} We use the normalization $b^{{\rm (n)}}=eB^{{\rm (n)}}_{\rm z}/m\omega_0 c$ ($B^{{\rm (n)}}_{\rm z}$ in the Gaussian units), and $\varkappa=x\omega_0/c$.  The normalized wave equation for $b^{{\rm (n)}}$ is:

\begin{equation}\label{Bwave}
\frac{\partial^{2} b^{{\rm (n)}}}{\partial \varkappa^{2}}-\frac{1}{\epsilon^{{\rm (n)}}} \frac{\partial \epsilon^{{\rm (n)}}}{\partial \varkappa} \frac{\partial b^{{\rm (n)}}}{\partial \varkappa}+n^2\left(\epsilon^{{\rm (n)}}-\sin ^{2} i\right) b^{{\rm (n)}}=f^{{\rm (n)}}
\end{equation}
where $f^{{\rm (n)}}$ is the source term, $f^{(1)}=0$ for first-harmonic and $f^{(2)}$ for second-harmonic is the normalized form of $\mathbf{F}^{(2\omega_0)}$ in Eq. (\ref{B2wave3}). The finite difference form of this equation is:

\begin{equation}\label{B2waveFD}
\begin{split}
\frac{b^{{\rm (n)}}_{\rm m+1}-2 b^{{\rm (n)}}_{\rm m}+b^{{\rm (n)}}_{\rm m-1}}{\Delta \varkappa^{2}}-\frac{\epsilon^{{\rm (n)}}_{\rm m+1}-\epsilon^{{\rm (n)}}_{\rm m-1}}{2 \Delta \varkappa \epsilon^{{\rm (n)}}_{\rm m}} \frac{\left(b^{{\rm (n)}}_{\rm m+1}-b^{{\rm (n)}}_{\rm m-1}\right)}{2 \Delta \varkappa}\\
+n^2\left(\epsilon^{{\rm (n)}}_{\rm m}-\sin ^{2} i\right) b^{{\rm (n)}}_{\rm m}=f^{{\rm (n)}}_{\rm m}
\end{split}
\end{equation}
One can cast Eq. (\ref{B2waveFD}) into tridiagonal form by grouping together the terms at common grid points. It gives:

\begin{equation}\label{tridiagonal}
\alpha^{\rm n}_{\rm m} b^{{\rm (n)}}_{\rm m-1}+\beta^{\rm n}_{\rm m} b^{{\rm (n)}}_{\rm m}+\gamma^{\rm n}_{\rm m} b^{{\rm (n)}}_{\rm m+1}=f^{\rm n}_{\rm m}
\end{equation}
where

\begin{subequations}\label{coeff}
\begin{align}
\alpha^{\rm n}_{\rm m} &=1+\frac{\Delta \epsilon^{{\rm (n)}}_{\rm m}}{\epsilon^{{\rm (n)}}_{\rm m}} \\
\beta^{\rm n}_{\rm m} &=n^2\left(\epsilon^{{\rm (n)}}_{\rm m}-\sin ^{2} i\right) \Delta \varkappa^{2}-2 \\
\gamma^{\rm n}_{\rm m} &=1-\frac{\Delta \epsilon^{{\rm (n)}}_{\rm m}}{\epsilon^{{\rm (n)}}_{\rm m}} \\
\Delta \epsilon^{{\rm (n)}}_{\rm m} &=\frac{\epsilon^{{\rm (n)}}_{\rm m+1}-\epsilon^{{\rm (n)}}_{\rm m-1}}{4} 
\end{align}
\end{subequations}

Equation (\ref{tridiagonal}) represents a tridiagonal system of equations which can be easily solved by the Thomas Algorithm or the Tridiagonal Matrix Algorithm (TDMA). The boundary condition for the first-harmonic is applied as below.  We suppose that $\varkappa_1=\Delta \varkappa$ lies in  vacuum region.  For $\varkappa_1$ we can split the wave into forward (input laser) and backward travelling waves with amplitudes $A_{0}$ and $R$, respectively.  Therefore, $b^{(1)}_{1}=A_{0}+R$. Using Eq. (\ref{tridiagonal}), we have

\begin{equation}\label{BC}
\alpha^{1}_{1} b^{(1)}_{0}+\beta^{1}_{1} b^{(1)}_{1}+\gamma^{1}_{1} b^{(1)}_{2}=0
\end{equation}
If the waves are moving at an angle $i$ relative to the $x-$axis,  direction of the density gradient, then the neighboring grid point will see a phase either advanced or receded by an amount $\Delta \phi=\varkappa\cos i$. Using $b^{(1)}\propto\exp(j \varkappa)$, we have

\begin{subequations}\label{coeff}
\begin{align}
b^{(1)}_{0} &=A_{0} e^{-j \Delta \varkappa \cos i}+R e^{j \Delta \varkappa \cos i} \\
&=b^{(1)}_{1} e^{j \Delta \varkappa \cos i}-2 j A_{0} \sin (\Delta \varkappa \cos i) 
\end{align}
\end{subequations}
Eliminating $b^{(1)}_{0}$ in Eq. (\ref{BC}) then gives 

\begin{equation}\label{BC2}
\kappa^{1}_{1} b^{(1)}_{1}+\gamma^{1}_{1} b^{(1)}_{2}=f^{1}_{1}
\end{equation}
where

\begin{subequations}\label{coeff}
\begin{align}
&f^{1}_{1}= 2 j A_{0} \sin (\Delta \varkappa \cos i) \\
&\kappa^{1}_{1}=\beta^{1}_{1}+e^{j \Delta \varkappa \cos i} 
\end{align}
\end{subequations}

This boundary condition can be also applied for the second-harmonic considering that second-harmonic is just in the reflected signal.


\section*{References}
\bibliography{manuscript}

\end{document}